
\magnification=\magstep1
\def\frac#1#2{{#1\over#2}}

\newcount\equationno      \equationno=0
\newtoks\chapterno \xdef\chapterno{}
\def\eqn{\eqno\eqname}
\def\eqname#1{\global \advance \equationno by 1 \relax
\xdef#1{{\noexpand{\rm}(\chapterno\number\equationno)}}#1}
\def\n{\noindent}
\def\la{\mathrel{\mathchoice {\vcenter{\offinterlineskip\halign{\hfil
$\displaystyle##$\hfil\cr<\cr\sim\cr}}}
{\vcenter{\offinterlineskip\halign{\hfil$\textstyle##$\hfil\cr<\cr\sim\cr}}}
{\vcenter{\offinterlineskip\halign{\hfil$\scriptstyle##$\hfil\cr<\cr\sim\cr}}}
{\vcenter{\offinterlineskip\halign{\hfil$\scriptscriptstyle##$\hfil\cr<\cr\sim\cr}}}}}

\def\ga{\mathrel{\mathchoice {\vcenter{\offinterlineskip\halign{\hfil
$\displaystyle##$\hfil\cr>\cr\sim\cr}}}
{\vcenter{\offinterlineskip\halign{\hfil$\textstyle##$\hfil\cr>\cr\sim\cr}}}
{\vcenter{\offinterlineskip\halign{\hfil$\scriptstyle##$\hfil\cr>\cr\sim\cr}}}
{\vcenter{\offinterlineskip\halign{\hfil$\scriptscriptstyle##$\hfil\cr>\cr\sim\cr}}}}}


\centerline{ \bf SCALING PROPERTIES OF GRAVITATIONAL CLUSTERING}
\centerline{\bf IN THE NON-LINEAR REGIME}
\vfill
\centerline{\bf R.Nityananda$^*$ and T.Padmanabhan$^\dagger$}
\vfill
\centerline{$^*$Raman Research Institute, Bangalore 560080, India}
\centerline{e-mail: rajaram@rri.ernet.in}
\medskip
\centerline{$^\dagger$Inter-University Centre for Astronomy and Astrophysics}
\centerline{Post Bag 4, Ganeshkhind, Pune 411 007, India}
\centerline{e-mail: paddy@iucaa.ernet.in}
\vfill
\centerline{ABSTRACT}
\noindent
The growth of density perturbations in an expanding universe in the
non-linear regime is investigated. The underlying equations of motion are
cast in a suggestive form, and motivate a conjecture that the scaled
pair velocity, $h(a,x)\equiv -[v/(\dot{a}x)]$ depends on the expansion
factor $a$ and comoving coordinate $x$ only through the density contrast
$\sigma(a,x)$. This leads to the result that the true, non-linear, density
contrast $<(\delta\rho/\rho)^{2}_{x}>^{1/2}=\sigma(a,x)$ is a universal
function of the density contrast $\sigma_L(a,l)$, computed in the linear
theory and evaluated at a scale $l$ where $l=x(1+\sigma^2)^{1/3}$. This
universality is supported by existing numerical simulations with
scale-invariant initial conditions having different power laws. We discuss
a physically motivated ansatz $h(a,x)=h[\sigma^2(a,x)]$ and use it to
compute the non-linear density contrast at any given scale analytically.
This provides a promising method for analysing the non-linear
evolution of density perturbations in the universe and for interpreting
numerical simulations.
\vfill
\centerline{IUCAA-12/93 April 93;Submitted for Publication}
\vfill\eject
\beginsection{1.Introduction}

It is generally believed that large scale structures in the universe
formed through the growth of small inhomogeneities via gravitational
instability. One convenient measure of the inhomogeneities present in the
{}~~universe ~at~~ any time $t$ is~ provided by the mean square~ fluctuation in
{}~the mass,~ $\sigma^2(t,x)=<(\delta M/M)^{2}_{x}>$, contained within a
sphere of comoving radius $x$. When $\sigma^2 \ll 1$, the evolution of
inhomogeneities can be studied using linear perturbation theory in an
expanding background. But when $\sigma^2$ grows to a value of order unity,
linear theory breaks down. This phase of the evolution is usually handled
by numerical simulations and analytic methods seem to have very limited
validity. Our understanding of structure formation will increase
significantly if the results of numerical simulations can be understood in
the framework of simple physical concepts and analytic approximations.
This paper discusses one possible approach to understand the evolution of
density fluctuation  in the nonlinear domain.

The idea behind this approach is as follows: It is known
(Peebles 1980) that the differential equation governing the correlation
function $\xi(t,x)$ is
essentially `driven' by the average pair velocity
$v_{\bf {pair}}(t,x)$. We present arguments which motivate the ansatz that the
ratio $[v_{\bf {pair}}(t,x)/-(\dot{a}x)]$ depends on $t$ and $x$ only
through the density contrast $\sigma^2(t,x)$; {\it i.e.,} we can set
$v_{\bf {pair}}(t,x)=[-\dot{a}xh(\sigma^2)]$ where $h(\sigma^2)$ is
some specific function which is independent of the initial conditions
chosen. With this physically motivated assumption, it is
possible to derive a differential equation for $\sigma^2$ and integrate it
in closed
form. The solution shows that, $\sigma^2(t,x)$ can be related to the
density contrast $\sigma^{2}_{L}(t,l)$ evaluated by linear theory at the
scale $l=x(1+\sigma^2)^{1/3}$ in a {\it universal} fashion. Such
universality, if established, would be a key feature of gravitational
dynamics in an expanding Universe. Using this result one can compute the
density contrast at highly non-linear epochs knowing the linear density
contrast evolved to the same epoch.

The paper is organized as follows: In the next section we discuss
Newtonian mechanics in an expanding universe and stress certain
properties of the resulting equations. In particular, as is well known,
they admit the possibility of similarity solutions in which the
statistical properties of density fluctuations at different times are
identical apart from a change of spatial scale. In section 3 we discuss
the similarity solution in some more detail and motivate possible
generalizations of this principle to estimate the non-linear density contrast.
The simplest
generalization does not work very well since it misses important
features of the gravitational dynamics. These
are discussed in section 4 which derives the main result of the
paper, estimates $\sigma^2$ at non-linear epochs and compares the
results with numerical simulations. The last section discusses the
implications and possibilities for extension.

\beginsection{2. Newtonian Gravity in expanding coordinates}

This section reviews some known aspects of clustering in a framework
somewhat different from, though equivalent to, the standard approach.
Consider a set of particles interacting via Newtonian gravity in an
expanding universe. We confine our attention to regions with
dimensions much smaller than the Hubble radius so that the equation of
motion for the $i^{th}$ particle is well approximated by:

$$
\ddot{\bf r}_i=- \sum_{j\neq i} \frac{Gm}{\mid{\bf r}_{ij}\mid^3}\,;\,\,\,{\bf
r}_{ij}={\bf r}_i-{\bf r}_j
\eqn\qq$$
Here ${\bf r}_i$ stands for the proper coordinate related to the comoving
coordinate ${\bf x}_i$ by ${\bf r}_i=a(t){\bf x}_i$. Using this we can
write the above equation as
$$
\ddot{\bf x}_i+\frac{2\dot{a}}{a}\dot{\bf x}_i=-\sum_{j\neq
i}\frac{Gm}{a^3}\frac{{\bf x}_{ij}}{\mid{\bf
x}_{ij}\mid^3}-\frac{\ddot{a}}{a}{\bf x}_i\equiv -\nabla\Phi
\eqn\two$$
The right hand side of \two\ has been expressed as the gradient of a scalar
potential $\Phi$ which obeys the equation
$$
\nabla^2\Phi =4\pi G\left[\sum_{i}\frac{m}{a^3}\delta({\bf x}-{\bf
x}_i)+\frac{3}{4\pi G}\left(\frac{\ddot{a}}{a}\right)\right]
\eqn\three$$
Equations \two\ and \three\ govern the dynamics of the particles in the
expanding background. In \three\ the ``mass density'' is contributed by two
terms, one of which is independent of {\bf x}. This fact can be used to
recast the equations in a more suggestive form by choosing $a(t)$ such
that it satisfies the  equation :
$$
\frac{\ddot{a}}{a}=-\frac{4\pi G\rho_0}{3}(\frac{a_0}{a})^3\equiv
-\frac{4\pi G}{3}\rho_b(t).
\eqn\qq$$
This equation  describes the  evolution of the expansion factor in the matter
dominated phase
of a Friedman universe.
Using these expressions we can write
\two\ and \three\ in the form
$$
\ddot{\bf x}_i+\frac{2\dot{a}}{a}{\bf \dot{x}}_i=-\nabla\Phi ;
\eqn\qq$$

$$
\nabla^2\Phi = \frac{4\pi G}{a^3}\left[\sum_{j}m\delta({\bf x}-{\bf
x}_j)-\rho_0a^{3}_{0}\right]
\eqn\qq$$
In the above form there is explicit dependence on $t$ due to the presence
of the terms $(\dot{a}/a)$ and $a^3$. It is useful to transform these
equations such that the time dependence disappears. Introducing new
dimensionless time and space coordinates $\tau$ and ${\bf q}_i$ via
$$
\tau \equiv \ln(t/T);\,\,\,{\bf x}_i\equiv L{\bf
q}_i;\,\,\,L^3=Gmt^{2}_{0}/a^{3}_{0}
\eqn\qq$$
with an arbitrary constant $T$, and transforming the equations, we easily find
that:
$$
\frac{d^2{\bf q}_i}{d\tau^2}+\frac{1}{3}\frac{d{\bf q}_i}{d\tau}=-\nabla_qU
\eqn\qq$$
$$
\nabla^{2}_{q}U=4\pi \sum_{i}\delta({\bf q}-{\bf q}_j)-\frac{2}{3}.
\eqn\qq$$

This equation has several noteworthy features which we shall briefly
comment upon.

(i) The compensating
negative background density, of $(-2/3)$ in the units chosen, cancels the long
range part of the gravity coming from the mean density. The usual
assumption made in the analysis of Jeans instability is thus justified
{\it in these coordinates}. One gets exponential growth in $\tau=\ln
t$ {\it i.e.,} power law growth in  $t$. More precisely, the analysis without
the
damping term would have led to modes  with $e^{\gamma\tau}$ with $\gamma^2
 =\frac{2}{3}$. With this term, one gets
$\gamma^2+\frac{1}{3}\gamma =\frac{2}{3}$, {\it i.e.,} $\gamma =\frac{2}{3}$ or
$-1$. The damping term has ensured that the growth and decay rates of the
two modes are {\it not} equal and one recovers the correct $t^{2/3}$
and $t^{-1}$  factors for the growing and decaying modes.

 (ii) The damping term is also
needed to recover other known results. For example, two bound bodies
spiral inwards with a shrinking period in the {\bf q} and $\tau$
coordinates, this being just an alternate description of a bound system
with a fixed size and period in {\bf r} and $t$.  It is also  instructive
to look at the free evolution of a test particle in an otherwise uniform
universe. Writing $p_i=dq_i/d\tau$, one gets $dp_i/d\tau = -\frac{1}{3}p_i$,
{\it i.e.,} $p_i\propto \exp (-\frac{1}{3}\tau )=t^{-1/3}$. This is
 the well known result that the peculiar velocity of a particle
decays as the inverse of the scale factor, when expressed in terms of $p$.

(iii) Particles at the boundary of a void feel the strong repulsion coming
from the locally uncompensated negative background term which pushes them
further out.

(iv) The translation invariance in $\tau$, {\it without
the damping term} would just lead to a conventional interacting system of
particles. The presence of the damping term reminds us that we should not
be trying to apply statistical mechanics with this interaction quite apart
from the well known difficulties associated with the singular short range
behaviour of Newtonian gravity.

(v) The usual calculation (multiplication
by $\dot{q}$ and integrating) which leads to
the  conservation of energy  for a time independent potential will now
show, because of the damping term, that the `energy' in these coordinates
always decreases, being constant only in the trivial
case in which all `velocities' and `accelerations' are zero. This decreasing
`energy' can be viewed as the driving force behind the continued evolution
with formation of bound structures and evacuation of voids.
 In fact, the form of behaviour closest to a steady state which is
known for this system is self similar clustering (Peebles 1980).  The
various distribution functions (two particle, three particle, etc.,),
describing the system have a particularly simple dependence on $\tau$.
Writing $dq/d\tau = p$, we can have for example a two particle
distribution function of the form

$$
f(q_1-q_2,\,p_1,
p_2;\tau)=e^{6\alpha\tau}g(e^{\alpha\tau}(q_1-q_2),e^{\alpha\tau}p_1,e^{\alpha\tau}p_2)
\eqn\qq$$
Here $f$ has been written as a function only of the coordinate difference
$q_1-q_2$ after averaging over the mean position $(q_1+q_2)/2$.
Notice that before such averaging the system is not self similar.

Finally, these equations suggest a possible way of interpreting
the success of adhesion models (Bagla and Padmanabhan, 1993 a). This
work is in progress and will be reported elsewhere.

\beginsection{3. Connecting linear and non-linear density contrast}

We  shall now describe the statistically self similar situation  in more
detail. Let us suppose that the power spectrum at some epoch has the form
$P(k)\propto k^n$, at sufficiently large scales which are linear. Then
$\sigma^{2}_{L}(x)\propto k^3P(k)\propto x^{-(3+n)}$ at the early, linear
stage.
Since $\sigma$ grows as $a$ in the linear theory, it follows that
$\sigma^{2}_{L}(x,a)\propto a^2x^{-(n+3)}$. This evolution is
self-similar with $\sigma^{2}_{L}\propto s^{-(n+3)}$ and
$s=(x/t^\alpha)$; $\alpha =[4/3(3+n)]$. In the extreme non-linear case,
bound structures with fixed proper radius $l=a(t)x$ will not participate
in the cosmic expansion. At a fixed $l,
\sigma^{2}_{NL}=<(\delta\rho/\rho_b)^2>$ must now grow as $(a^6/a^3)=a^3$.
The $a^6$ factor arises from the fact that the background density $\rho_b$
decreases as $a^{-3}$ and $\sigma^2\propto\rho^{-2}_{b}$; the $a^3$ factor
in the denominator arises from the fact that in all samples of proper
radius $l$ used in computing the variance, only a fraction proportional to
$a^{-3}$ will contribute. Thus, in the non-linear regime
$\sigma^{2}_{NL}(t,x)\propto a^3(t) F(a(t)x)$. The form of $F$ can be
determined by matching the linear and non-linear expressions at say,
$\sigma=\sigma_c$. This will lead to the standard result (Peebles 1980) that
$\sigma^{2}_{NL}\propto
a^3[ax]^{-\gamma}$ with $\gamma = 3(n+3)/(n+5)$. Thus we find that

$$
\sigma^{2}_{L}(a,x)\propto a^2(t)x^{-(n+3)};\,\,\,\sigma^{2}_{NL}(a,x)\propto
a^3[ax]^{-\gamma}\propto a^{6/(n+5)}x^{-3(n+3)/(n+5)}
\eqn\qq$$
A graphical version of this argument is given in Fig.1.
Self-similarity of this solution is evident from the fact that
$\sigma_{NL}\propto s^{-\gamma}$. Notice that, in the range $n>-2$, the
non-linear correlation function at fixed comoving scale $x$ grows more
{\it slowly} compared to the linear spectrum. It follows from these
relations that one can write

$$\eqalign{
\left(\frac{\sigma_{NL}}{\sigma_c}\right) &=
\left(\frac{\sigma_L}{\sigma_c}\right)^{3/(n+5)}\,\,\,\,({\rm
for}\,\,\sigma_L\gg \sigma_c) \cr
 &= \left(\frac{\sigma_L}{\sigma_c}\right)\,\,\,\,({\rm for}\,\,\sigma_L\ll
\sigma_c),
\cr}\eqn\fstnl$$
where $\sigma_c$ is the typical value around which the transition to
non-linearity  occurs. We shall now examine this result in
greater detail.

The above result \fstnl\ is supposed to give the non-linear density contrast at
a
given scale provided the linear density contrast at the same scale is
given. Unfortunately, this result is of very little practical utility and
suffers from several limitations as it stands. Let us study these limitations
since
they need to be tackled in any attempt to improve upon the relation.

To begin with, this result is obtained by assuming a power law form for
the spectrum. The power spectrum {\it e.g.,} just after recombination, is
not a pure power law in any of the realistic models of structure
formation. Hence it is not clear what is the value of $n$ which one should
use. One could, of course, try to provide a local definition of $n$ by
taking $n_{\bf {eff}}=(d\ln P/d\ln k)$. Unfortunately, the resulting
$\sigma_{NL}$ does not lead to even rough agreement with the numerical
simulations.

Secondly, notice that the {\it actual} value of $\sigma_{NL}$ depends on
the value of $\sigma_c$ chosen. It is difficult to estimate $\sigma_c$
from theoretical considerations with any degree of accuracy. Even if
$\sigma_c$ is estimated, the above relation can be valid only for
$\sigma_{NL} \gg \sigma_c $ and $\sigma_{NL} \ll \sigma_c .$ In the regime with
$\sigma_{NL} \simeq \sigma_c$ we do not have a reliable estimate.

Finally, notice that the above expression attempts to relate
$\sigma_{NL}(x,a)$ to $\sigma_L$ at the same $x$ and $a$. This is somewhat
unrealistic because non-linear growth at a given scale cannot be
expected to be a strictly local function of the linear density contrast at
the {\it same} comoving scale.

Let us now ask how some of the above difficulties can be circumvented by a
more sophisticated approach. One possibility which suggests itself is to
write a differential equation for $\sigma_{NL}$ which can be generalized
to the case in which the power spectrum is not a pure power law. This is
fairly easy to do by noting that

$$
\frac{\partial\ln\sigma}{\partial\ln
a}=\frac{3}{(n+5)};\,\,\,\frac{\partial\ln\sigma}{\partial\ln
x}=\frac{3(n+3)}{2(n+5)}
\eqn\qq$$
Hence, eliminating $n$,

$$
\frac{\partial\ln\sigma}{\partial\ln a}-\frac{\partial\ln
\sigma}{\partial\ln x}=\frac{2}{3}.
\eqn\secnl$$
Such an equation implies that at each epoch and scale, one is evolving $\sigma$
using the result of the self similar solution for the currently and
locally applicable conditions. This equation is assumed to hold for
$\sigma > \sigma_c$. For $\sigma <
\sigma_c$ we use the rather trivial evolution equation of the linear theory:

$$
\frac{\partial\ln\sigma}{\partial\ln a}=1.
\eqn\qq$$
These equations are to be interpreted as follows. At some very early
epochs (say, at $a=a_{\bf {rec}}$) we are given the density contrast
$\sigma(x,a_{\bf {rec}}(x))$. We will assume that the initial epoch is
chosen sufficiently early so that $\sigma_{\bf {rec}}\ll 1$ at all relevant
scales. With this initial condition the above equations can be integrated
forward for all further times thereby giving $\sigma(a,x)$. This approach
clearly does not require $\sigma$ to be a power law.

Because of the extreme simplicity, the evolution equation above can be
integrated in a general form. The general solution for $\sigma > \sigma_c$
given by

$$
\ln\sigma_{NL}(a,x)=\frac{3}{2}\ln\left(\frac{a}{a_{\bf {rec}}}\right)+F(ax)
\eqn\qq$$
where $F$ is a function to be determined by matching $\sigma_{NL}$ and
$\sigma_L$ at $\sigma = \sigma_c$. Since the linear density contrast
evolves as $\sigma_L \propto a$, it follows that we can relate the
$\sigma_{NL}$ and $\sigma_L$ as follows:

$$
\sigma_{NL}(a,x)=\sigma^{3/2}_{L}(a,l);\,\,\,l=x\sigma^{2/3}_{NL}(a,x)\,\,\,\,({\rm
for}\,\,\sigma_{NL} > \sigma_c)
\eqn\qq$$
This shows that when $\sigma > \sigma_c, \sigma_{NL}$ at a point $x$ is
determined by $\sigma_L$ at a point $l=x\sigma^{2/3}_{NL}$. In other words
non-linear evolution does introduce a degree of nonlocality in comoving
scale $x$ when this approximation is used.

While the above attempt is an improvement, it is still not very
satisfactory. The differential equation \secnl\ is difficult to derive
directly from any quantitative dynamical considerations. It is merely the
simplest equation that can be written down in which the power law index
$n$ is eliminated. That this equation is inadequate is clear from two
aspects of our solution. Firstly, it does not make a smooth transition
from linear to non-linear scales and secondly it does not provide us with
a value for $\sigma_c$. Nevertheless the solution suggests a very simple
ansatz for the evolution of the density contrast: It may be possible to
take the true density contrast, $\sigma_{NL}(a,x)$, to be a universal
function of $\sigma_L[a,f(x,\sigma_{NL})]$ where $f(x,\sigma_{NL})\simeq
x$ for $\sigma_{NL}\ll 1$ and $f(x,\sigma_{NL})\simeq x\sigma^{2/3}_{NL}$
for $\sigma_{NL}\gg 1$. [In fact, a relation with this structure has
recently been suggested by Hamilton et al. (1991) and used to obtain the
input linear spectrum from the observed non-linear structure observed at
the current cosmological epoch. Our emphasis in this paper is on exploring
the physical origin,  dynamical derivation and implications of such a
scaling relation.] In such a more realistic theory, we expect $f$ to vary
smoothly from one limit to the other. It turns out that such a scaling
relation can indeed to be obtained from the equations governing the
evolution of density perturbations. This is task which we shall now
take up.

\beginsection{4. Non-linear dynamics - a better approximation}

We begin by defining the correlation function $\xi(x,t)$ as
the fourier transform of the power spectrum:

$$
\xi({\bf x},t)=\int \frac{d^3{\bf k}}{(2\pi )^3}P(k,t)e^{{\bf
ik.x}}=\int^{\infty}_{0}\frac{dk}{k} \Delta^2(k,t)\left(\frac{\sin
kx}{kx}\right)
\eqn\qq$$
To an excellent approximation, $\xi(x,t)$ and $\sigma (x,t)$ are related
to each other by
$$
\sigma^2(x,t)\cong\frac{3}{x^3}\int^{x}_{0}\xi(y,t)y^2 dy
\eqn\corsig$$
We shall now obtain an {\it exact} equation satisfied by $\sigma(a,x)$ by
an argument which follows the treatment of (Peebles 1980) for
the evolution of the correlation function $\xi$. Since the mean number of
neighbours to any
particle is given by

$$
N(x,t)=(na^3)\int^{x}_{o}4\pi y^2dy[1+\xi(y,t)]
\eqn\qq$$
when $n$ is the comoving number density, the conservation law for pairs
implies

$$
\frac{\partial\xi}{\partial t}+\frac{1}{ax^2}\frac{\partial}{\partial
x}[x^2(1+\xi)v]=0
\eqn\corev$$
where $v$ denotes the mean relative velocity of pairs at scale $x$ and
epoch $t$. Using \corsig\ , we find

$$
(1+\xi)=\frac{1}{3x^2}\frac{\partial}{\partial x}[x^3(1+\sigma^2)]
\eqn\qq$$
Substituting this in \corev\ , we get

$$
\frac{1}{3x^2}\frac{\partial}{\partial x}[x^3\frac{\partial}{\partial
t}(1+\sigma^2)] = -\frac{1}{ax^2}\frac{\partial}{\partial
x}\left[\frac{v}{3} \frac{\partial}{\partial x}[x^2(1+\sigma^2)]\right]
\eqn\qq$$
Integrating, we find:

$$
x^3 \frac{\partial}{\partial
t}(1+\sigma^2)=-\frac{v}{a}\frac{\partial}{\partial x}[x^3(1+\sigma^2)]
\eqn\qq$$
The integration would allow the addition of an arbitrary function of $t$ on
the right hand side. We have set this function to zero so as to reproduce
the correct limiting behaviour (see below).
It is now convenient to change the variables from $t$ to $a$, thereby
getting an equation for $\sigma^2$:

$$
a\frac{\partial}{\partial
a}[1+\sigma^2(a,x)]=\left(\frac{v}{-\dot{a}x}\right)
\frac{1}{x^2}\frac{\partial}{\partial x}[x^3(1+\sigma^2(a,x))]
\eqn\twentysix$$
This equation shows that the behaviour of $\sigma^2(a,x)$ is essentially
decided by the dimensionless ratio $h(a,x)\equiv[v(a,x)/(-\dot{a}x)]$
between the mean relative velocity $v$ and the Hubble velocity
$\dot{a}x=(\dot{a}/a)x_{\bf {prop}}$. To understand the behaviour of this
equation let us consider its solutions in two limiting cases. In the
non-linear limit, peculiar motion exactly compensates for the Hubble
expansion to form bound structure; hence $(v/-\dot{a}x)=1$ giving,

$$
a\frac{\partial}{\partial a}(1+\sigma^2)-x\frac{\partial}{\partial
x}(1+\sigma^2)=3(1+\sigma^2)
\eqn\qq$$
This equation has the general solution $(1+\sigma^2)=a^3F(ax)$. In the
linear limit, we know that $\sigma^2=Aa^2x^{-(n+3)} \ll 1$, if the power
spectrum is $P(k)\propto k^n$. Ignoring $\sigma^2$ compared to unity on
the right hand side of \twentysix, we get

$$
a\frac{\partial}{\partial a}[Aa^2x^{-(n+3)}]\cong
3\left(\frac{v}{-\dot{a}x}\right) = 2Aa^2x^{-(n+3)}
\eqn\qq$$
Therefore,

$$
\left(\frac{v}{-\dot{a}x}\right)=(2/3)Aa^2x^{-(n+3)}=(2/3)\sigma^2\propto
[t^{-4/3(n+3)}x]^{-(n+3)}\propto s^{-(n+3)}
\eqn\qq$$
where $s=(x/t^\alpha)$ and $\alpha=[4/3(n+3)]$. The solution in the
non-linear limit, $(1+\sigma^2)=a^3F(ax)$ will be a function of $s$ alone
only if $F(ax)=(ax)^\gamma$ with $\gamma=6/(3\alpha +2)$; in that case
$a^{3-\gamma}x^{-\gamma}\propto s^{-6/(3\alpha +2)}$. This recovers the
`standard results' and justifies the boundary conditions chosen earlier.
Also note that, in the linear limit, $(v/-\dot{a}x)=(2/3)\sigma^2$. This
result in the linear regime is true for any spectrum, not for just a power law
(Peebles 1980).

It is now clear that any general relation between
$\sigma_{NL}$ and $\sigma_L$ will translate itself into a similar relation
for $h$ in the two limits.  Conversely, the behaviour of $h(a,x)$ will
allow us to determine $\sigma_{NL}$ in terms of $\sigma_{L}$. The
results obtained above suggest the hypothesis that the quantity
$h(a,x)=[v(x,a)/-\dot{a}x]$ is purely a function of $\sigma^2$.

$$
h(a,x) = h[\sigma^2(x)]
\eqn\qq$$
In the linear limit, when $\sigma^2 \ll 1$, we saw that $h(a,x)\cong
(2/3)\sigma^2$; in the extreme non-linear limit $(\sigma^2 \gg 1)$ bound
structures would have formed in which $v$ will balance out Hubble
expansion; that is, $v=-\dot{a}x$ implying $h\cong 1$. It seems reasonable
to explore the assumption that $h$ depends on $a$ and $x$ only through
some universal function $h=h(\sigma^2)$ which has the asymptotic behaviour
derived above. Of course, this assumption has to be ultimately checked by
comparing the results with the numerical  simulations [ A word of caution on
our
notation is in order. We are using the same symbol $h$ to denote two
functional forms $h(a, x) $ and $ h (\sigma^2 )$. A more pedandic
notation will be $h(a, x) = H(\sigma^2)$, introducing another symbol $H$.
We will not do this since it should be clear from the context what we mean by
$h$ ].

When $h(a,x)=h[\sigma^2(a,x)]$, it is possible to find a useful solution
to \twentysix.
We are interested in the solution to this equation which reduces
to the form $\sigma^2 \propto a^2 x^{-(n+3)}$ for $\sigma^2 \ll 1$. This
can be obtained as follows: We rewrite the equation in the form

$$ a {\partial D \over \partial a} - h(D) x{\partial D \over \partial x} = 3h
D;
\eqn\baq $$
where $D = (1+\sigma^2)$. Suppose we can find a function $F(D)$ such that
$$F(D) = \alpha \ln a + \beta \ln x \eqn\qdeff $$
with $\alpha, \beta$, being constants. If this is true, then
$$ F' dD = {\alpha \over a} da + { \beta \over x} dx \eqn\qq $$
giving $ a(\partial D / \partial a) = (\alpha / F')$ and
$x(\partial D / \partial x) = (\beta / F')$. Substituting this
into \baq, we find that $F(D)$ must satisfy the relation

$$F'(D) = { \alpha - \beta h \over 3hD } \eqn\qq$$

\noindent Or, equivalently,

$$ F = \int \left( {\alpha - \beta h \over 3hD } \right) dD = \int {( \alpha -
\beta h) \over 3h} { d \sigma^2 \over (1 + \sigma^2 )} \eqn\qans $$

\noindent The solution $\sigma^2 = \sigma^2 (a, x)$ is obtained by
combining \qdeff\ and \qans\.  We can easily find that
$$\int {d\sigma^2\over h(\sigma^2)(1+\sigma^2)}=\ln
\left[a^3\left\{x^3(1+\sigma^2)\right\}^m\right]\eqn\qbsol$$
where $m \equiv (\beta /\alpha)$ is a constant. When $h(\sigma^2)\simeq
(2/3)\sigma^2$
and $\sigma^2\ll 1$, we can ignore $\sigma^2$ compared to $\ln \sigma^2$
and this relation gives the linear theory result:
$\sigma^2_L\propto a^2 x^{2m}$ allowing us to relate $m$ to the index $n$ of
the
power spectrum. Since we expect $\sigma^2_L \propto a^2 x^{-(n+3)}$ if
$P(k) \propto k^n$, we get
 $m=-(n+3)/2$ . This is the solution with the correct boundary condition which
we need. It is also easy to verify that when
$h\simeq 1$, $\sigma^2\gg 1$, we get
$$\sigma^2\propto (a^3 x^{3m})^{1/(1-m)}
\propto a^3(ax)^{3m/(1-m)}
\propto a^3 (ax)^{-\gamma}\eqn\qq$$
where $\gamma=3m/(m-1)$
$=3(n+3)/(n+5)$. Thus the limiting forms
derived earlier arise from the  limiting behaviour of $h(\sigma^2)$.

The solution we have found also exhibits the universality we noticed in the
earlier approximations. To see this,
note that the right hand side of \qbsol\ is a function
of $\sigma^2_L(a,l=x(1+\sigma^2)^{1/3})$ $\propto a^2 x^{2m}$
$(1+\sigma^2)^{2m/3}$, which is
the density contrast $\sigma_L(a,l)$
in the linear theory {\it evaluated at} the scale $l=x(1+\sigma^2)^{1/3}$.
Thus we can write
$$\sigma^2_L[a,l]=\exp {2\over 3}
\int {d\sigma^2\over h(\sigma^2)(1+\sigma^2)}\eqn\qsilex$$
Since the original power law
index has disappeared, this relation suggests
that $\sigma^2_L(a,l)$ is a universal function of the correct
$\sigma^2(a,x)$. This is precisely the kind of ansatz suggested by our
earlier approximation. The function $f(x, \sigma) = x(1 + \sigma^2)^{1/3}$
has the limiting form, $f \simeq x \sigma^{2/3}$ for $\sigma \gg 1 , and f
\simeq
x $ for $ \sigma \ll 1$.

The above relation between $\sigma_L^2$ and $\sigma^2$ is indeed the correct
solution to equation \baq\ even when $\sigma_L^2$ is not a power law. This
can be seen by direct substitution of the solution \qsilex\ into \baq\ or --
more formally -- by the follwoing argument. Let $A = \ln a, X = \ln  x$ and
$D(X, A)$ be the solution to \baq\ .
We define curves in the $(X, A)$ such that
$$ { dX \over dA}\big\vert_c = -h(D [x, A] ) \eqn\qq $$
i.e. the tangent to the curve at any point $(X, A)$ has the value
determined by $h$ at that point. Quite clearly, along the curve,
the left hand side of \baq\ is a total derivative allowing us to write
$$ \left ( {\partial D \over \partial A} - h(D) {\partial D \over \partial X}
\right)_c = \left( {\partial D \over \partial A} + {\partial D \over \partial
X}
{dX \over dA } \right)_c = { dD \over dA}\big\vert_c = 3hD \eqn\qq $$
Integrating
$$ \exp { 1 \over 3} \int\limits_c { dD \over Dh(D) } = \exp A = a \eqn\qq $$
Or, Equivalently,

$$ \exp {2 \over 3} \int\limits_c { d\sigma^2 \over h(\sigma^2) (1 + \sigma^2 )
}
= a^2 \propto \sigma_L^2 \big\vert_l \eqn\qq $$
We not only need to determine the form of the curves to fix the scale $l$.
The equation to the curve can now be written as

$$ {dX \over dA} = - h = - { 1 \over 3D} { dD \over dA} \eqn\qq $$

\n giving

$$ 3X + \ln D = \ln \left[x^3 (1 + \sigma^2 ) \right] = {\rm constant} \eqn\qq
$$
This shows that $\sigma_L^2$ should be evaluated at fixed
$ l = x(1 + \sigma^2)^{1/3}$, giving us the result of \qsilex\ .

The actual relation between $\sigma_{NL}$ and $\sigma_{L}$ depends on the forms
of
$h(\sigma^2)$.  From the limiting behaviour discussed before, we know that $h
(\sigma^2) \simeq (2/3) \sigma^2$ for small $\sigma$ and
$h \simeq 1$ for $\sigma \gg 1$.  Our results for the nonlinear density
contrast
depends crucially on the manner in which $h(\sigma)$ reaches unity.  We will
expect
an overdense region to expand more slowly compared to background,
reach a maximum radius, collapse and  (vivialise) to form a bound structure.
During the collapse phase the  average
velocity will (in general)  overshoot the Hubble expansion velocity. Only after
varialization is complete will the pair correlation velocity approach the
asymptotic
value of $(-\dot ax )$. In fact numerical simulations   show
that $h(\sigma^2)$ has a single maximum (and hence overshoots
$|v|=\dot a x$ before falling back) at about $\sigma^2\simeq (8-15)$
with $h_{max}\simeq (1.5-2)$. The simplest model for such a
function with correct asymptotic behaviour will be:
$$h(\sigma^2)={2\over 3}
\sigma^2
{(1+\lambda \sigma^2)\over
(1+(2/3)\lambda \sigma^4)}\eqn\qq$$
which has just  one free parameter $\lambda$. Substituting this in \qsilex
and integrating, we find that
$$\sigma^2_L(a,l)=\sigma^2(1+\lambda \sigma^2)^{(3\lambda+2)/3(1-\lambda)}
(1+\sigma^2)^{-(3+2\lambda)/3(1-\lambda)}\eqn\qq$$
where $\sigma^2=\sigma^2(a,x)$. Best agreement with the
numerical results is achieved for $\lambda\simeq 0.36$ for which
$$\sigma^2_L(a,l)=\sigma^2(1+0.36\sigma^2)^{1.604}
(1+\sigma^2)^{-1.937} ; l = x(1 + \sigma^2)^{1/3}\eqn\qour$$

\n This result gives the true $\sigma^2(a,x)$ as a function of the linear
density
contrast $\sigma^2_L (a,l)$ evaluated at $l = x(1 + \sigma^2)^{1/3}.$

This result does not suffer from the limitations of the earlier
approximations. It gives the exact density contrast $\sigma^2$ in
terms of the density contrast of the linear theory $\sigma_L^2$ by a function
which
makes a smooth transition from $\sigma^2 \ll 1$ regime to
$\sigma^2 \gg 1$ regime. The non locality of behaviour seen in our earlier
approximations is also preserved in a smooth manner.

The ultimate test of any such
formula, of course, is based on the agreement with numerical simulations. It
turns out
that our formula does remarkably well on this count. To begin with, the results
of the numerical simulations with scale invariant power spectra does show the
kind of
universality obtained above: The correct density contrast $\sigma (a, x)$
is indeed expressible as a universal function of the linear density contrast
$\sigma_L(a, l)$ where $l = x (1 + \sigma^2)^{1/3}$. The best fitting function
to the
numerical data is the multiparameter fit.
$$\sigma^2_L=\sigma^2
\left[{1+0.0158\sigma^4+0.000115\sigma^2\over
1+0.926\sigma^4-0.0743\sigma^6+0.0156\sigma^8}\right]^{1/3}\eqn\qham$$
In figure 2 we have plotted this multiparameter fit as well as our result in
equation
(32).It is remarkable that our single parameter fit is reasonable
for all $\sigma^2\ga 6$ and for $\sigma^2\la 0.2$. The deviation
in the intermediate range is due to the fact that \qour\ has the behaviour
$(\sigma_L/\sigma)^2$ $=1+{\cal O}(\sigma^2)$ while the actual
data suggests $(\sigma_L/\sigma)^2$
$=1+{\cal O}(\sigma^4)$ for $\sigma^2_L\ga 10$.

It should be stressed that $\sigma^2_L$ and $\sigma^2$ in the above
formulas are evaluated at different length scales. The
relation of $\sigma^2$ to $\sigma^2_L$ at the {\it same} length scale is
quite complicated and depends on the shape of the spectrum. For
example, in the extreme nonlinear limit,
$$\sigma^2(x,a)\simeq
11.4^{2/(n+5)} a^{6/(n+5)}(x/x_0)^{-\gamma} \simeq
11.4^{2/(n+5)}
a^{-2(n+2)/(n+5)}
\sigma^2_L (x,a)(x/x_0)^{\beta}\eqn\qq$$
with $\beta=[(n+3)(n+2)]/(n+5)$,
$\gamma= 3(n+3)/(n+5)$
 and
$x_0\simeq 8h^{-1}Mpc$.
For a CDM spectrum, galactic scales have $n\simeq -2$
and smaller scales have $n\simeq -3$. So the nonlinear
density contrast is a factor of about $11$ larger at subgalactic scales
and a factor of $11^{2/3}$ higher at galactic scales. This has
the effect of steepening the correlation function at the nonlinear end.

\beginsection{5. Discussion and Conclusions}

The basic proposal of this paper is that significant progress in
understanding the evolution of the correlation function in the non-linear
phase of gravitational clustering can be made via a scaling hypothesis
. This assumption - that the pair velocity is a function only of
the density contrast on the same scale - leads to the explicit formulae of
section 4.

It will be interesting to test the hypothesis on a wider range
of initial conditions than the power law cases explored so far. Given that
both the velocity and the density contrast are averages over a variety of
environments, the emergence of such a relation is not obvious (except in
the extreme linear and non-linear regimes) and deserves dynamical
validation. It would be interesting to see whether the form of
the function $h (\sigma^2 ) $ can be obtained from some approximate
dynamical models
{\it e.g.,} the Zeldovich approximation, the  frozen flow model etc.
In a widr context one would also like to investigate the conditions
under which the
 universality is  broken . For
example, one can visualise two scenarios which start with the same input
power spectrum, one having random phases (gaussian fluctuations) and the
other with non-gaussian correlations. One does not, in general, expect the
resulting correlation functions to match in the non-linear regime. Even in
such situations, analysis in terms of deviations from `universal' scaling
might be an economical and physical way of compressing the information
from a large ensemble of simulations. These issues are under investigation
(Bagla and
Padmanabhan, 1993b).
\medskip
\centerline{\bf Acknowledgement}
\medskip
\n This work was initiated when one of the authors (R. Nityananda) was visiting
the Theoretical Astrophysics Group, TIFR, Bombay.
\beginsection \centerline{\bf REFERENCES}

\item{} Bagla, JS and T. Padmanabhan (1993a), Nonlinear evolution of density
perturbation using approximate constancy of gravitational potential,
IUCAA-.../93
preprint.

\medskip

\item{} Bagla, JS and T. Padmanabhan (1993b) work in progress.

\medskip

\item{} Efsthathiou, G et al., (1988) Mon. Not. R. Astron. Soc., {\bf 235},
715.

\medskip

\item{} Hamilton, A.J.S et al., (1991), Astrophys. J. {\bf 374} L1 - L4.
\medskip
\item{} Peebles, P.J.E. (1980) {\bf The Large Scale Structure of the
Universe}, (Princeton University Press, Princeton).
\vfill\eject

\n{\bf Figure Captions}

\medskip

\n {\bf Fig. 1} A schematic diagram showing the  geometrical origin
of Peebles scaling rules discussed in the text. The correlation function is
plotted against the proper length in a logarithmic scale. The two curves
indicate the
function $\xi (ax)$ at two different epochs $a = a_1$ and at $a = a_1e$.
At large scales,
in which the evaluation  is linear, the correlation function changes through
two processes.
The scale is
stretched due to expansion and the linear density   evolves as
$\delta \propto a$ . At very small scales, where the evaluation is non-linear,
the
length scale does not change due to expansion, but the correlation function
should increase as  the cube of the expansion factor . Elementary geometry now
shows
that if the slope of the line at the linear end is $\left[-(n + 3)\right]$ then
the slope
at the non-linear end must be $ \left[{3 (n + 3) \over (n + 5) }\right]$.
For a more detailed
discussion see text.

\medskip

\n {\bf Fig. 2} The true density contrast $\sigma^2$ is plotted against the
linear density contrast. Note that the two contrasts are calculated at two
different
scales as discussed in the text. The unbroken curve is an exact fit to the
numerical simulations while the broken curve gives the result of this paper.
There is fair agreement between the analytic approximation and
numerical simulations for density contrasts of the order of 10 and above.
\end

\end